\documentclass[twocolumn,showpacs,preprintnumbers,amsmath,amssymb]{revtex4}

\usepackage{graphicx}
\usepackage{dcolumn}
\usepackage{bm}

\begin{document}

\preprint{Tsujii et al}

\title{Universality in heavy-fermion systems with general degeneracy}

\author{N. Tsujii$^1$ \email{TSUJII.Naohito@nims.go.jp}, H. Kontani$^2$, and K. Yoshimura$^3$}
\affiliation{$^1$National Institute for Materials Science,
 Sengen 1-2-1, Tsukuba, 305-0047, Japan}
\affiliation{$^2$Graduate School of Science, Nagoya University, Fro-cho, Chikusa-ku, Nagoya City 464-8602, Japan}
\affiliation{$^3$Graduate School of Science, Kyoto University, Kyoto 606-8502, Japan}

\date{\today}

\begin{abstract}%
We discuss the relation between the $T^2$-coefficient of electrical resistivity $A$ 
and the $T$-linear specific-heat coefficient $\gamma$ for heavy-fermion systems with general $N$, 
where $N$ is the degeneracy of quasi-particles. A set of experimental data reveals that 
the Kadowaki-Woods relation; $A/\gamma^2$ = $1\times10^{-5} {\rm \mu\Omega}$(K mol/mJ)$^2$, 
collapses remarkably for large-$N$ systems, 
although this relation has been regarded to be commonly applicable to the Fermi-liquids. 
Instead, based on the Fermi-liquid theory we propose a new relation; 
$\tilde{A}/\tilde{\gamma}^2=1\times10^{-5}$ 
with $\tilde{A}$ = $A/\frac{1}{2}N(N-1)$ and
$\tilde{\gamma} = \gamma/\frac{1}{2}N(N-1)$. 
This new relation exhibits an excellent agreement with the data for 
whole the range of degenerate heavy-fermions.
\end{abstract}

\maketitle

The Fermi-liquid theory~\cite{Landau} is the most fundamental one to understand 
the electronic state of metallic systems. 
This theory has achieved a great success in describing not only 
the electronic properties of normal metals but also unusual properties 
of the strongly-correlated electron systems~\cite{Yamada,Yanase} 
like $f$-electron based heavy-fermion compounds~\cite{Yamada,Yanase,Stewart,Brandt} 
and $d$-electron based intermetallics and oxides ~\cite{Yamada,Moriya}.
In this theory, the effect of electron-electron interactions are involved 
in the effective mass of quasi-particles, $m^*$. 
This enables a very simple representation of physical properties: 
the electronic specific heat $C$ and the electrical resistivity $\rho$ are described 
as $C = \gamma{}T$ and $\rho = AT^2$ with $\gamma \propto m^*$ and $A \propto m^{*2}$. 
Such a temperature dependence has been actually observed in numerous kind of metals. 
Moreover, this description implies that the ratio $A/\gamma^2$ does not depend on $m^*$, 
resulting in the universal value of $A/\gamma^2$. 
In fact, it has been revealed that many $f$-electron based systems show 
the universal behavior; 
$A/\gamma^2$ = 1.0$\times$10$^{-5}$ ${\rm \mu\Omega{}cm(K mol/mJ)^2}$ ~\cite{Kadowaki}, 
called as the Kadowaki-Woods (KW) relation. 
The KW-relation has therefore been accepted as the most essential relation 
showing the validity of the Fermi-liquid theory.

Recently, however, significant and systematic deviations from the relation 
have been observed in many heavy-fermion compounds, in spite that 
they apparently show Fermi-liquid behavior at low temperature ~\cite{Tsujii}. 
This class of compounds includes Yb-based compounds like 
YbCu$_5$, YbAgCu$_4$, YbCuAl, YbNi$_2$Ge$_2$, YbInCu$_4$, YbAl$_3$, 
and Ce-based compounds like CeNi$_9$Si$_4$~\cite{Michor} and CeSn$_3$. 
Notably, the deviations in these systems are almost 'universal'; 
$A/\gamma^2 \approx 0.4\times10^{-6} {\rm \mu\Omega{}cm(K mol/mJ)^2}$. 
This systematic and large deviation cannot be explained by specific characters of materials, 
like carrier density, band structure, anisotropy, etc.
Instead, there seems to exist a common physical origin. 
The origin of this deviation therefore rises an important issue 
for the generality of the Fermi-liquid theory.

Very recently, a theoretical work based on the Fermi-liquid theory has suggested ~\cite{Kontani} 
that the values of $A/\gamma^2$, so far considered to be unique and independent on materials, 
do depend on the number of degeneracy of quasi-particles $N$. 
For isolated atoms, $N$ is defined as $N = 2J +1$ with $J$ the total angular momentum. 
In solids, $N$ can vary due to the competition between the crystal-field splitting $\Delta$ 
and the Kondo temperature $T_{\rm K}$. 
For $T_{\rm K} < \Delta$, the low-temperature properties are basically 
explained by $N$ = 2 ($S$ = 1/2) Kondo model, 
since most of the degeneracy are lost due to the large $\Delta$ ~\cite{Brandt,Besnus,Fujita}. 
For $T_{\rm K} > \Delta$, on the contrary, the crystal field splittings are 
covered by the large Kondo effect, 
and the degeneracy are almost preserved down to low temperatures. 
In this case, the theory ~\cite{Kontani} gives the failed universality of $A/\gamma^2$.

In this paper, we make a quantitative comparison of the experimental data in ref.\cite{Tsujii} and
several recent works with the theoretical results of ref.\cite{Kontani}. 
The results display a beautiful agreement between experiments and theory. 
Furthermore, we propose an advanced relation for $A$ and $\gamma$ based on these results. 
Using $\tilde{A}$ and $\tilde{\gamma}$, the values of $A$ and $\gamma$ normalized by $\frac{1}{2}N(N-1)$, 
we show that these two values of heavy-fermion systems with general $N$ are 
related by a very simple formula; $\tilde{A}/\tilde{\gamma}^2$ = 1$\times$10$^{-5}$ 
${\rm \mu\Omega cm(K mol/mJ)^2}$. 
This new relation, namely, the `grand-KW-relation', will be an important waymark
for the research of strongly-correlated electron systems with degeneracies, 
and remarkably extends the validity of the Fermi-liquid theory.

At first, we briefly describe the theoretical results of ref.~\cite{Kontani}. 
For the case of strong-coupling limit where $m^*/m \gg 1$ 
($m^*$ and $m$ being the mass of heavy quasi-particles and free electrons, respectively), 
the orbitally-degenerate periodic Anderson (ODPA) model gives~\cite{Comment0}:
\begin{eqnarray}
 A &=& \frac{hk_{\rm B}^2}{e^2}\frac{3\pi^6}{2k_{\rm F}^{4}a^3}N(N-1){\rm \Gamma_{loc}^2}(0,0){\rm \rho_{f}^4}(0),\\
 \gamma &=& N_{\rm A}k_{\rm B}^{2}\frac{\pi^2}{6}N(N-1){\rm \Gamma_{loc}}(0,0){\rm \rho_{f}^2}(0) .
\end{eqnarray}
where $h$ is the Plank constant, $k_{\rm B}$ the Boltzmann constant, $k_{\rm F}$ the Fermi momentum, 
and $N_{\rm A}$ the Avogadro number. In addition, $a$ is the unit-cell length, and ${\rm \rho_{f}(0)}$ 
the density of states per $f$-orbit at the Fermi energy. 
$\Gamma_{\rm loc}$(0,0) represents the effective interaction between quasi-particles. 
Note that $A$ and $\gamma$ given in eq.(1) and (2) are not simply proportional to $N(N - 1)$, 
because $\Gamma_{\rm loc}$(0,0) also depends on $N$. 
The value $A/\gamma^2$ is then deduced as ~\cite{Comment1}:
\begin{eqnarray}
  \frac{A}{\gamma^2} &=& \frac{h}{e^{2}k_{\rm B}^{2}N_{\rm A}^{2}}\cdot\frac{9(3\pi^2)^{-1/3}}{n^{4/3}a^3}\frac{1}{\frac{1}{2}N(N-1)} \nonumber \\
                     &\approx& \frac{1\times10^{-5}}{\frac{1}{2}N(N-1)} \hspace*{3mm}{\rm \mu\Omega{}cm(K mol/mJ)^2} .
\end{eqnarray}
For the case of $N$ = 2, this formula gives the KW-relation. 
For general $N$, this gives a set of universal relations. 
This is shown in Figure 1 as the solid lines for $N$ = 2, 4, 6, and 8.
\begin{figure}[tb]
\begin{center}
\includegraphics[width=8cm]{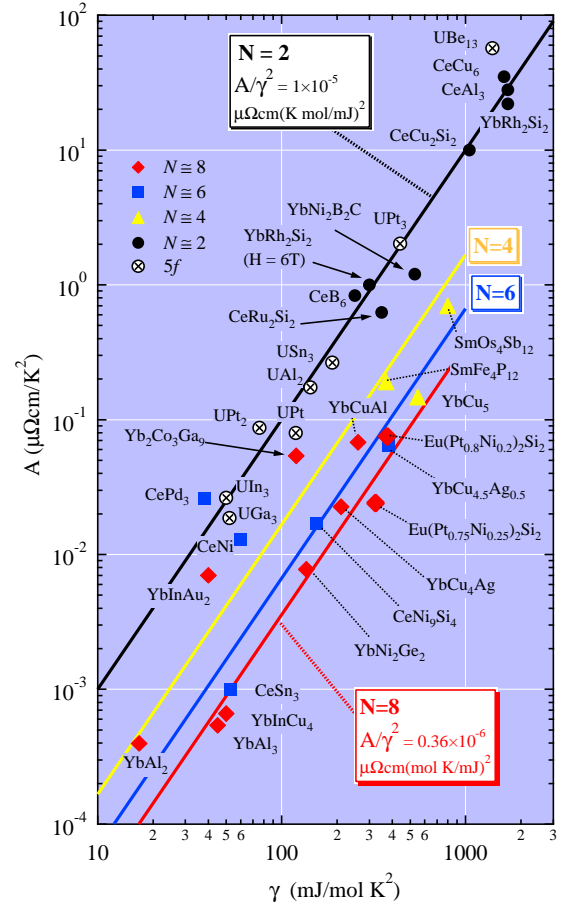}
\end{center}
\caption{
$T^2$-coefficient of electrical resistivity $A$, vs. $T$-linear coefficient of specific heat $\gamma$ 
of heavy-fermion systems with various degeneracy. 
Experimental data are taken from ref.\cite{Kadowaki,Tsujii,Michor,Mitsuda,Takeda,Sanada}. 
The black line corresponds to the Kadowaki-Woods relation~\cite{Kadowaki}.
Other solid lines are the prediction from the orbitally-degenerate periodic-Anderson model~\cite{Kontani}. 
Colors of the symbols represent the degeneracy $N$ probable for the systems; 
black, yellow, blue, and red indicate $N$ = 2, 4, 6 and 8, respectively. 
The value of $N$ of U-compounds are not determined.
}
\end{figure}

In the figure, experimental data are also plotted after 
ref.\cite{Kadowaki,Tsujii,Michor,Mitsuda,Takeda,Sanada,Comment2}.
At first, one can see that many heavy-fermion systems such as CeCu$_6$ or CeCu$_2$Si$_2$
agree with the KW-relation; i.e., the theoretical prediction for $N=2$.
This is consistent with the situation $T_{\rm K} < \Delta$, which results in
the low degeneracy of $N=2$ ~\cite{CeB6}.
Moreover, it is clear that many Yb- and Ce-based systems, which have deviated from the KW-relation, 
well agree with the theoretical predictions for $N$ = 6-8. 

It should also be noted that the $A/\gamma^2$ of Eu- and Sm-based 
compounds agrees very well with the line for $N$ = 8 and $N$ = 4, respectively. 
These Eu-compounds are considered to be intermediate-valent between Eu$^{2+}$($S$ = 7/2) 
and Eu$^{3+}$($J$ = 0)~\cite{Mitsuda}.
The Fermi-liquid state of them is hence considered to be emerged 
out of the degeneracy $N = 2S +1 = 8$.
For the two Sm-based systems, the value of $N$ = 4 has been assumed, 
since the lowest CEF levels are considered to be a quartet~\cite{Takeda,Sanada}.
These quantitative agreements of $A/\gamma^2$ with respective theoretical lines 
evidences that $A/\gamma^2$ of heavy-fermion systems are not specific to materials,
but are commonly scaled by degeneracy.

In the following, we go forward to unify these relations into a single relation. 
If the value of $N$ is determined experimentally, we can define the normalized coefficients 
$\tilde{A}$ and $\tilde{\gamma}$ from the eq.(1) and (2) as:
\begin{equation*}
 \tilde{A}=\frac{A}{\frac{1}{2}N(N-1)}, \hspace*{2mm} \tilde{\gamma}=\frac{\gamma}{\frac{1}{2}N(N-1)} .
\end{equation*}
Then $\tilde{A}/\tilde{\gamma}^2$ is obtained from the eq.(3) as:
\begin{equation}
\tilde{A}/\tilde{\gamma}^2 \approx 1\times10^{-5} \hspace*{3mm}{\rm \mu\Omega{}cm(K mol/mJ)^2} .
\end{equation}
This formula does not include any $N$-dependence. 
Hence, this should be applicable to arbitrary-$N$ systems.

In Figure 2, we plot $\tilde{A}$ and $\tilde{\gamma}$ of $f$-electron based systems.
\begin{figure}[tb]
\begin{center}
\includegraphics[width=8.5cm]{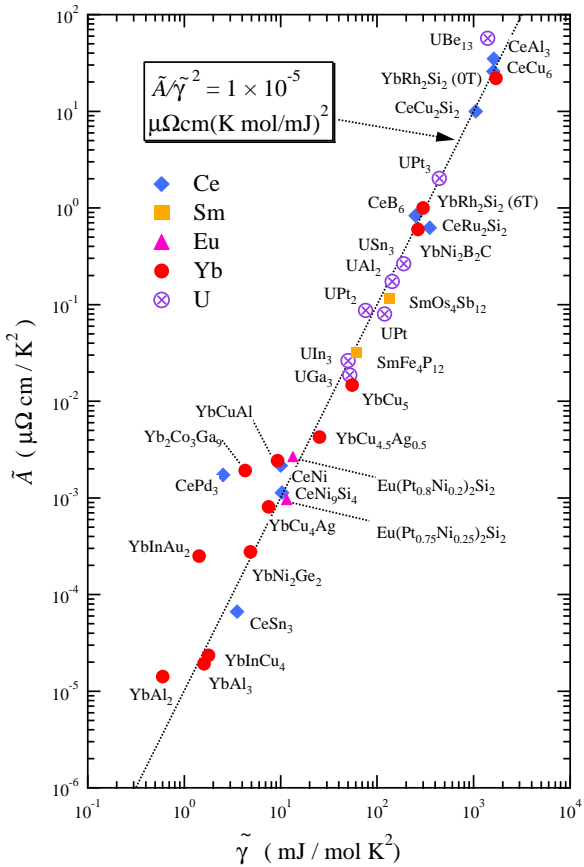}
\end{center}
\caption{
The plot of $\tilde{A}$ and $\tilde{\gamma}$ of heavy-fermion systems. 
$\tilde{A}$ and $\tilde{\gamma}$ are the divided values of $A$ and $\gamma$ by $\frac{1}{2}N(N - 1)$, respectively. 
$N$ of the U-based compounds is tentatively assumed to be 2. 
The dotted line represents the grand KW-relation (4) given in the text; 
$\tilde{A}/\tilde{\gamma}^2$ = 1$\times$10$^{-5}$ ${\rm \mu\Omega}$(K mol/mJ)$^2$.
}
\end{figure}
For uranium compounds, we have tentatively assumed $N$ = 2, 
which is discussed later. 
One can see that the eq.(4) scales $\tilde{A}$ and $\tilde{\gamma}$ universally 
for a wide range of materials. 
This fact shows the validity of the theoretical approach using the ODPA model, 
and extends the validity of the Fermi-liquid theory. 
We would like to stress that our new relation has the same form 
just as the original KW-relation. 
The formula (4) may hence be called as `grand-KW-relation for general degeneracy'.

Here it would be interesting to discuss to what extent this rule holds
when system is reached toward the quantum critical point (QCP).
Even in the vicinity of the QCP, the Fermi-liquid state is realized
at sufficiently low temperatures below a characteristic temperature
($T_{\rm coh}$, in literatures) as far as the system is in the magnetically
disordered side.
In this case, our theory yields that the value of $\tilde{A}/\tilde{\gamma}^2$
defined at the low temperature limit ($T \rightarrow 0$),
follows the relation (4) even in the vicinity of QCP.
This is because our theory is derived for the limit of $T \rightarrow 0$.
It should also be noticeable to point that the theoretical calculations
of $A/\gamma^2$ for $N$ = 2 based on the spin-fluctuation theory show
that the ratio is approximately independent of the distance from the 
QCP~\cite{Takimoto,Continentino}.
In fact, $A/\gamma^2$ of YbRh$_2$Si$_2$ and CeInCo$_5$,
both of which are considered to be in the vicinity of QCP, are almost a constant
as external magnetic field is varied~\cite{Gegenwart,Bianchi}.
Meanwhile, for the case of `very' close to the QCP, where $T_{\rm coh}$ is quite low,
a deviation from the universality may be observed, as is suggested 
theoretically~\cite{Takimoto} and experimentally on YbRh$_2$(Si,Ge)$_2$~\cite{Custers}.
This deviation however seems to occur in an extremely narrow condition
where the Fermi-liquid description is probably not valid.
Except for such extreme cases, the grand-KW relation is one of the common behavior
of Fermi liquids, even in systems close to the QCP.
Note that the $f$-orbital degeneracy will stabilize the Fermi-liquid state,
because $N$-dependence of $T_{\rm K}$ ($\propto e^{-1/\rho NJ_{\rm K}}$)
will be much prominent than that of $T_{\rm N}$($\propto N^2J_{\rm RKKY}$).
A large mass-enhancement is realized with relatively higher $T_{\rm K}$
when $N > 2$.

There should, of course, exist exceptions.
As is seen in the formula (3),
the ratio $A/\gamma^2$ as well as that of $\tilde{A}/\tilde{\gamma}^2$ 
depend on the carrier concentration $n$, 
wave number at the Fermi energy $k_{\rm F}$, and so on. 
If one of these values are extremely different from typical ones, 
$\tilde{A}/\tilde{\gamma}^2$ can deviate remarkably.
Such an example is CePd$_3$. 
In Fig.2, one can see that CePd$_3$ shows a large deviation from eq.(4),
though CePd$_3$ well agrees with the original KW-relation~\cite{Kadowaki}. 
This discrepancy results from the large degeneracy, $N$ = 6 for CePd$_3$~\cite{Murani}. 
It should be noted that CePd$_3$ has very small carrier-concentration 
(0.3 electrons per f.u.)~\cite{Webb}. 
The $A/\gamma^2$ value is found to depend on $n$ as proportional to $n^{-4/3}$
from eq.(3) and also from other theoretical studies
~\cite{Yamada,Takimoto,Continentino,carrier}.
Taking this into consideration, 
the deviation of CePd$_3$ from the universal line is reasonable.
Similar deviation is reported for the Kondo semiconductor CeNiSn~\cite{Terashima}.
Anomalously large $A$ value (54 ${\rm \mu\Omega{}cm/K^2}$) compared to its 
$\gamma$ (40 mJ/mol K$^2$) has been attributed to its extremely low carrier concentration
~\cite{Terashima}.

The compounds CeNi($N$ = 6) and YbCuAl($N$ = 8) also show slight deviations, 
possibly due to the error in the $N$ estimations. 
For other exceptions, YbInAu$_2$ and Yb$_2$Co$_3$Ga$_9$, 
we have no explanation for the origin of deviation. 
Other causes such as multi-Fermi-surface effect~\cite{Michor} may have to be considered. 
In addition, strong anisotropy of the Fermi surface can cause 
deviation from the universal relation~\cite{Yamada,Fukazawa}.
This effect would be in general more prominent in $d$-electron systems~\cite{Hussey,Comment3}.

For U-based compounds, its degeneracy has been the subject of arguments. 
If the 5$f$-electrons are well localized, $N$ can be determined experimentally, 
as in the case of UPd$_3$ ~\cite{Buyers}. 
In most of U-compounds, however, it is considered that the 5$f$-electrons have 
more-itinerant character than 4$f$, since 5$f$-orbitals are spacially more expanded. 
The definition of $N$ in U-compounds is therefore ambiguous. 
Here, one can see in Fig.1 and Fig.2 that those U-compounds well agree 
with the theoretical prediction for $N$ = 2. 
This can lead us to the possibility that the orbital degree of freedom is quenched 
and only the spin degree of freedom participates in the Fermi-liquid state in these 5$f$-systems, 
similar to transition metals. 
Although the estimation of $N$ from the $A/\gamma^2$ plot is not conclusive, 
this plot may serve as a hint to discuss the puzzling 5$f$-electrons.

In addition, we note that the grand-KW-relation is also powerful to describe 
the pressure dependent properties of heavy-fermion systems. 
In CeCu$_2$Ge$_2$ (or YbNi$_2$Ge$_2$), it is suggested that 
the value of $A/\gamma^2$ reduces (or increases) 
about 25 times at high pressures~\cite{Jaccard,Knebel}
probably due to the change of $N$ by pressures.
In our plot of $\tilde{A}$ and $\tilde{\gamma}$, these crossover would be described 
on the single scaling without breaking the universality. 
This situation may be hence ideal for the continuity principle of the Landau Fermi-liquid theory~\cite{Anderson}. 
Current interests in strongly-correlated electron systems are extended 
to the orbitally-degenerate cases.
Hence, the grand-KW-relation will be one of the most fundamental relation in Fermi-liquid systems. 
We also comment that the effect of the degeneracy must be taken into consideration 
in many other physical quanta, like anomalous Hall effect~\cite{Hall}, etc. 
The analysis using the ODPA model will be henceforth indispensable.

Authors acknowledge K. Yamada, A. Mitsuda, Y. Aoki, G. Kido and H. Kitazawa
for fruitful discussions and comments.


\begin{thebibliography}{99}
\bibitem{Comment0} The eq.(1) and (2) explicitly describes the units like k$_{\rm B}$ and $h$,
that were omitted in the equations of ref.~\cite{Kontani}.
\bibitem{Comment1} Here the free electron model $k_{\rm F}$ = (3$\pi^2{}n$)$^{1/3}$ 
($n$ being the carrier concentration) is employed. 
For numerical estimation, we have assumed $n^{4/3}a^{3} \approx 1\times10^{8}$cm$^{-1}$.
This assumption corresponds to 3.3 carriers per formula unit for $a$ = 5\AA{}
or to 2.8 carriers per f.u. for $a$ = 4\AA{},
which seems to be typical values for them.
Formula unit is to include only one rare-earth ion.
\bibitem{Comment2} For the values of $N$, we have employed the data listed in ref.~\cite{Tsujii}, 
most of which were determined by fitting the specific heat using the impurity-Kondo models. 
Although the real system is periodic, we believe that this estimation of $N$ is reliable, 
because these fittings are performed for the data at relatively high temperatures, 
at which intersite effect are not prominent. 
However, it is difficult to determine the value of $N$ explicitly when $\Delta$
(crystal-field splitting) and $T_{\rm K}$ are comparable.
\end{thebibliography}
\end{document}